\documentclass{article}

\usepackage{amsmath,amssymb,amsfonts}   
\usepackage{graphicx,bm}

\newtheorem{theorem}{\bf Theorem}[section]

\renewcommand{\L}{{\mathbb L}}

\newcommand{\e}{{\mathrm e}}

\newcommand{\calG}{{\mathcal G}}
\newcommand{\calA}{{\mathcal A}}
\newcommand{\calC}{{\mathcal C}}

\newcommand{\Markov}[2]{\underset{#1}{\overset{#2}{\rightleftharpoons}}}

\begin{document}

\title{\bf Multi-spike solutions of a hybrid reaction-transport model}

\author{ \em
P. C. Bressloff, \\ Department of Mathematics, 
University of Utah \\155 South 1400 East, Salt Lake City, UT 84112}

 \maketitle

\begin{abstract}
Numerical simulations of classical pattern forming reaction-diffusion systems indicate that they often operate 
in the strongly nonlinear regime, with the final steady-state consisting of a spatially repeating pattern of localized spikes. In activator-inhibitor systems such as the two-component Gierer-Meinhardt (GM) model, one can consider the singular limit $D_a\ll D_h$, where $D_a$ and $D_h$ are the diffusivities of the activator and inhibitor, respectively. Asymptotic analysis can then be used to analyze the existence and linear stability of multi-spike solutions.
In this paper, we analyze multi-spike solutions in a hybrid reaction-transport model, consisting of a slowly diffusing activator and an actively transported inhibitor that switches at a rate $\alpha$ between right-moving and left-moving velocity states. This class of model was recently introduced to account for the formation and homeostatic regulation of synaptic puncta during larval development in {\em C. elegans}. We exploit the fact that that the hybrid model can be mapped onto the classical GM model in the fast switching limit $\alpha\rightarrow \infty$, which allows us to establish the existence of multi-spike solutions. Linearization about the multi-spike solution leads to a non-local eigenvalue problem that is used to derive stability conditions for the multi-spike solution for finite $\alpha$. \end{abstract}

\section{Introduction}

One major mechanism for self-organization between (and within cells) is the combination of diffusion and nonlinear chemical reactions. The notion that a reaction-diffusion (RD) system can spontaneously generate spatiotemporal patterns was first introduced by Turing in his seminal 1952 paper on morphogenesis \cite{Turing52}. He established the principle that two nonlinearly interacting chemical species differing significantly in their rates of diffusion can amplify spatially periodic fluctuations in their concentrations, resulting in the formation of a stable periodic pattern. The Turing mechanism for morphogenesis was subsequently refined by Gierer and Meinhardt \cite{Gierer72}, who showed that one way to generate a Turing instability is to have an antagonistic pair of molecular species known as an activator-inhibitor system, which consists of a slowly diffusing chemical activator and a quickly diffusing chemical inhibitor. For many years the only direct experimental evidence for spatiotemporal patterning of molecular concentrations came from the inorganic Belousov-Zhabotinsky reaction \cite{Zaikin70}, until Kondo and Asai demonstrated the occurrence of the Turing mechanism in studies of animal coat patterning \cite{Kondo95}. More recent advances in live cell imaging and gene knockout protocols are now allowing for a closer connection between theories of pattern formation and developmental cell biology. The range of models and applications of the Turing mechanism have expanded dramatically \cite{Murray08,Cross09,Walgraef97}, including applications to ecology \cite{Levin92,Rietkerk07} and systems neuroscience \cite{Ermentrout79,Bressloff01}; in the latter case nonlocal synaptic interactions drive pattern forming instabilities rather than diffusion. 

There are also a growing number of examples of spatiotemporal patterns of signaling molecules at the intracellular level \cite{Kholodenko09,Howard12}. Such patterns typically regulate downstream structures such as the cytoskeleton and cell membrane, which then drive various mechanical processes including cell division and polarization \cite{Goryachev08,Drake10}. As highlighted by a number of authors \cite{Ishihara07,Otsuji07,Halatek18,Halatek18a,Brauns20}, the classical activator-inhibitor mechanism for Turing pattern formation in cell development does not apply to many of the known examples of intracellular patterning during cell polarization and division. First, these latter processes are mass-conserving, at least on the time-scales for patterns to initially form and stabilize. Second, rather than two or more chemical species diffusing in the same medium and mutually affecting their rates of production and degradation, intracellular patterns typically involve the dynamical exchange of proteins between the cytoplasm and plasma membrane, resulting in associated changes of conformational state and a spatial redistribution of mass.  One example of intracellular pattern formation without mass conservation is an RD model of the self-positioning of structural maintenance of chromosomes (SMC) protein complexes in {\em E. coli}, which are required for correct chromosome condensation, organization and segregation \cite{Murray17,Murray19,Murray20}. An interesting features shared by all of these model systems is that they tend to operate in the strongly non-linear regime (far from pattern onset). In the mass-conserving case, the final steady-state often consists of a single localized peak (spike) of the slowly diffusing activator, whereas multi-spike solutions can exist when mass is not conserved \cite{Murray20}. That is, the initial Turing pattern predicted from linear theory undergoes significant coarsening.

\begin{figure}[t!]
\begin{center} \includegraphics[width=12cm]{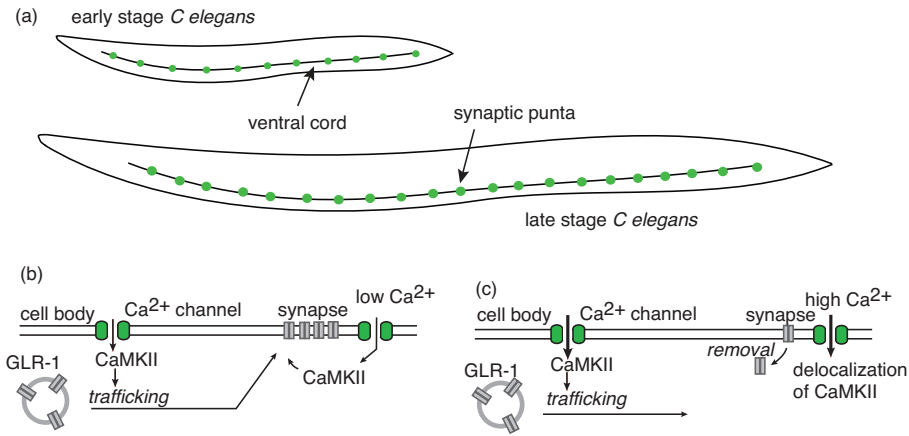} \end{center}
\caption{(a) Schematic figure showing distribution of synaptic punta along ventral cord of early and late stage {\em C. elegans}. New synapses are inserted during development in order to maintain the synaptic density. (b) Regulation of transport and delivery of GLR-1 to synapses by CaMKII. Calcium influx through voltage-gated calcium channels activates CaMKII, which enhances the active transport and delivery of GLR-1 to synapses. Under conditions of increased excitation, higher calcium levels results in constitutively active CaMKII which fails to localize at synapses, leading to the removal of GLR-1 from synapses.}
\label{fig1}
\end{figure}

Based on experimental studies of synaptogenesis in \textit{Caenorhabditis elegans}, we recently introduced a hybrid reaction-transport model for intracellular pattern formation, which involves the interaction between a passively diffusing component and an advecting component that switches between anterograde and retrograde motor-driven transport (bidirectional transport) \cite{Brooks16,Brooks17}. These components were identified as the protein kinase CaMKII and the glutamate receptor GLR-1, respectively. During larval development of {\em C. elegans}, the density of ventral and dorsal cord synapses containing the glutamate receptor GLR-1 is maintained despite significant changes in neurite length \cite{Rongo99}, see Fig. \ref{fig1}(a). It is known that the coupling of synapse number to neurite length requires CaMKII and voltage-gated calcium channels, and that CaMKII regulates the active (kinesin-based) transport and delivery of GLR-1 to synapses \cite{Shen99,Monteiro12,Hoerndli13,Hoerndli15}, see Fig. \ref{fig1}(b). However, a long outstanding problem has been  identifying a possible physical mechanism involving diffusing CaMKII molecules and motor-driven GLR-1 that leads to the homeostatic control of synaptic density. Using the classical Gierer and Meinhardt mechanism for reaction kinetics \cite{Gierer72}, we showed that our model supported Turing patterns on a one-dimensional domain of fixed length. (Turing pattern formation based on advecting species has also been considered within the context of chemotaxis \cite{Hillen96}.) Numerical simulations of the model using experimentally-based parameters generated patterns with a wavelength consistent with the synaptic spacing found in {\em C. elegans}, after identifying the in-phase CaMKII/GLR-1 concentration peaks as sites of new synapses. Extending the model to the case of  a slowly growing 1D compartment, we subsequently showed how the synaptic density can be maintained during {\em C. elegans} growth, due to the insertion of new concentration peaks as the length of the compartment increases \cite{Brooks17}. 

The above example of synaptogenesis in {\em C. elegans} motivates developing a more general theory of pattern formation in hybrid reaction-transport systems, particularly given the prevalence of both active and passive transport within cells. As part of such a program, we focus in this paper on multi-spike solutions far from pattern onset. Our previous work used linear stability analysis to identify regions in parameter space for a Turing instability close to pattern onset. However, numerical simulations indicated that the hybrid reaction-transport system operated in the strongly nonlinear regime, with the final steady-state consisting of a spatially repeating pattern of localized spikes.
Although weakly nonlinear analysis is not applicable in the large amplitude regime, it is still possible to study the existence and stability of spike patterns in certain limits. For example, in the case of classical RD systems such as the Brusselator, Gierer-Meinhardt and Gray-Scott models, one can consider the singular limit $D_a\ll D_h$, where $D_a$ and $D_h$ are the diffusivities of a slowly diffusing activator and a fast diffusing inhibitor, say. These systems support multi-spike solutions in which the activator is localized to narrow peak regions (spikes) within which the inhibitor is slowly varying. Asymptotic analysis can then be used to analyze the existence and linear stability of multi-spike solutions \cite{Iron01,Iron02,Wei03,Kol05a,Kol05b,Tzou13}, see also \cite{Murray20}. 

The structure of the paper is as follows. The hybrid reaction-transport model of \cite{Brooks16,Brooks17} is introduced in section 2, which is a three-component model consisting of a passively diffusing activator concentration $P$ and a pair of actively transported inhibitor concentrations $A_{\pm}$, which correspond to left-moving and right-moving velocity states, respectively. In steady-state, the full model is shown to be equivalent to the classical two-component GM model for $(P,A)$, where $A=A_++A_-$. In particular, the effective diffusivity of the inhibitor takes the form $D=v^2/(\mu_2+2\alpha)$, where $v$ is the speed of each velocity state, $\alpha$ is the switching rate, and $\mu_2$ is the degradation rate. In section 3 we construct a multi-spike solution of the steady-state model using the asymptotic method introduced in \cite{Iron01}. The one major difference is that in addition to the concentrations $P,A$, one also has to determine the effective flux $J=v(A_+-A_-)$. In section 4 we turn to the linear stability analysis of the multi-spike solution. This requires taking into account perturbations of all three components of the hybrid model, which ultimately leads to a non-local eigenvalue problem that is a generalization of the one obtained in \cite{Wei03}. Nevertheless, one can adapt the asymptotic analysis and winding number arguments of \cite{Wei03} to derive conditions for the stability of a multi-spike solution. In particular, we show how stability depends on the switching rate and degradation rate of the inhibitor.

\section{Model}


Consider a one-dimensional domain of fixed length $2L$.  Let $P(x,t)$ denote the concentration of a passively transported component at position $x$ at time $t$ and let $A(x,t)$ denote the corresponding actively transported component. The latter is partitioned into two subpopulations: those that undergo anterograde transport ($R_+$) and those that undergo retrograde transport ($R_-$) with
$A(x,t)=A_+(x,t)+A_-(x,t)$.
Individual particles randomly switch between the two advective states according to a two-state Markov process
$A_+ \Markov{\alpha}{\alpha}A_-$,
with transition rate $\alpha$. Modeling the interactions between the active and passive using the classical activator-inhibitor system of Gierer and Meinhardt (GM) \cite{Gierer72}, we then have the following system of equations \cite{Brooks16}:
\begin{subequations}
\label{RDmodel0}
\begin{align}
\frac{\partial P}{\partial t}&=D_P\frac{\partial^2 P}{\partial x^2}+  f(P,A_++A_-) , \\
\frac{\partial A_+}{\partial t}&=-v\frac{\partial A_+}{\partial x}+\alpha A_- -\alpha A_++ g(P,A_+) ,\quad x\in -[-L,L],\\
\frac{\partial A_-}{\partial t}&=v\frac{\partial A_-}{\partial x}+\alpha A_+-\alpha A_-+  g(P,A_-) ,
\end{align}
\end{subequations}
where $D_P$ is the diffusion coefficient of the passive component and $v$ is the speed of the motor-driven active component. 
The GM reaction terms are
\begin{equation}
f(P,A)=\rho_1 \frac{P^2}{A} -\mu_1P,\quad g(P,A)=\rho_2 P^2-\mu_2A,
\end{equation}
so that $P$ is the activator and $A_{\pm}$ the inhibitor. Here
 $\rho_1, \rho_2$ represent the strength of interactions, and $\mu_1,\mu_2$ are degradation rates. Equations (\ref{RDmodel}) are supplemented by reflecting boundary conditions at the ends $x=\pm L$:
\begin{align}
&\left . \frac{\partial C(x,t)}{\partial x}\right |_{x=\pm L}=0,\quad vR_+(\pm L,t)=vR_-(\pm L,t).
\end{align}
(Note that in the specific application to synaptogenesis in {\em C. elegans}, $P$ represented CaMKII and $A_{\pm}$ represented GLR-1 \cite{Brooks16,Brooks17}.)

Adding and subtracting the pair of equations (\ref{RDmodel0}b,c) and setting $A=A_++A_-$ and $J=vA_+-vA_-$ yields the equivalent system
\begin{subequations}
\label{RDmodel}
\begin{align}
\frac{\partial P}{\partial t}&=D_P\frac{\partial^2 P}{\partial x^2}+ \rho_1 \frac{P^2}{A} -\mu_1P , \\
\frac{\partial A}{\partial t}&=-\frac{\partial J}{\partial x}+2\rho_2 P^2-\mu_2A,\quad x\in -[-L,L],\\
\frac{\partial J}{\partial t}&=-{v^2} \frac{\partial A}{\partial x}-(2\alpha +\mu_2)J.
\end{align}
\end{subequations}
Analogous to the reduction of a bidirectional transport model to the telegrapher's equation \cite{Kac74}, introducing the effective diffusivity
\begin{equation}
D=\frac{v^2}{2\alpha+\mu_2},
\end{equation}
and taking the limit $\alpha\rightarrow \infty$ and $v^2\rightarrow \infty$ with $D$ fixed  (so that we can set $\partial HJ/\partial t=0$) yields the classical GM model for $(P,A)$ \cite{Gierer72}
\begin{subequations}
\label{RD00}
\begin{align}
\frac{\partial P}{\partial t}&=D_P\frac{d^2 P}{d x^2}+    \frac{\rho_1P^2}{A} -\mu_1P, \\
\frac{\partial A}{\partial t}&=D\frac{d^2A}{dx^2}+2\rho_2 P^2-\mu_2A ,\quad D=\frac{ v^2}{2\alpha+\mu_2},
\end{align}
\end{subequations}
In particular, we can identify $D$ as the diffusivity of the inhibitor in the reduced model. In addition, the time-independent versions of equations (\ref{RDmodel}) and (\ref{RD00}) are also equivalent for all finite values of $\alpha,D$. It follows that the three-component hybrid reaction-transport model and the two-component RD model  have the same steady-state solutions for $(P,A)$. This includes the unique uniform solution 
\begin{equation}
P^*=\frac{\rho_1\mu_2}{2\rho_2\mu_1},\quad A_{\pm}^*=\frac{\rho_1^2\mu_2}{4\rho_2\mu_1^2},
\end{equation}
and multi-spike solutions. However, the corresponding eigenvalue problem obtained by linearizing about a given steady-state solution will differ in the two models, which means that their stability properties could also differ. A related issue is that, given a spatially varying steady-state solution for $(P,A)$ in the RD model, the corresponding hybrid system could support several different solutions for the flux $J(x)$ and thus $A_{\pm}(x)$. In summary, one cannot simply carry over previous results for the existence and stability of spike solutions in the GM model of pattern formation \cite{Iron01}. 

One necessary condition for the occurrence of pattern formation in activator-inhibitor systems such as the GM model is that the inhibitor diffuses more quickly than the activator. Moreover, in order to obtain spike solutions of the form considered in Refs. \cite{Iron01,Iron02}, the diffusivities should differ from each other by several orders of magnitude. The above analysis suggests that $D$ is the effective diffusivity of the actively transported component. We thus obtain the condition $D_P\ll D$. In the particular application to synaptogenesis in {\em C. elegans} the following order of magnitude parameter values were used \cite{Brooks16,Brooks17}: $\mu_1 ,\mu_2,\rho_1 \sim 1$/s, $\rho_2\sim 1/(\mu M\cdot $s), $\alpha \sim 0.1$/s, $v \sim 1\mu\ $m/s and $D_P\sim 0.01\ \mu$m$^2$/s. In addition, the typical spacing of synaptic puncta is 3-4 per 10 $\mu$m. It is clear that for this application, the effective diffusivity $D\approx 1\ \mu$m$^2$/s, and thus the inequality $D_P\ll D$ is satisfied. In this paper, we fix the units of time and space by setting $\mu_1=1$ and $L=1$, and take $\rho_1=2\rho_2=1$. (In the case of {\em C. elegans} this corresponds to units $t=1$s and $x=10\ \mu$m, assuming that $L$ is identified with a section of ventral cord containing approximately 3 or 4 synaptic puncta. An adult has a length of around 1 mm.) Writing $D_P=\epsilon^2$ with $0<\epsilon \ll 1$, we then have three remaining parameters $\mu_2,\alpha,D$. It is also more convenient to work with the variables $P,A$ and $J$. Performing the rescaling $P\rightarrow  {\epsilon}  P/\mu_2$, $A\rightarrow  {\epsilon}  A/\mu_2$, $J\rightarrow \epsilon J/\mu_2^2$ and $v^2\rightarrow \mu_2 v^2$ then gives
\begin{subequations}
\label{RDmodel0}
\begin{align}
\frac{\partial P}{\partial t}&=\epsilon^2 \frac{\partial^2 P}{\partial x^2}+   \frac{P^2}{A} - P , \\
\frac{1}{\mu_2}\frac{\partial A}{\partial t}&=-\frac{\partial J}{\partial x}+\frac{1}{\epsilon} P^2-A,\quad x\in [-1,1] ,\\
\frac{1}{2\alpha+\mu_2} \frac{\partial J}{\partial t}&=-D\frac{\partial A}{\partial x}-J.
\end{align}
\end{subequations}

\section{Existence of an $N$-spike solution}

We are interested in deriving conditions for the existence of multiple spike solutions in the limit $\epsilon \rightarrow 0$. This type of solution consists of localized peaks in $P$ of width $\epsilon $ with $P$ exponentially small outside each spike, while $A_{\pm}$ change slowly within each spike. Here we extend the asymptotic analysis of the classical GM model \cite{Iron01,Iron02}. Setting time-derivatives to zero in equations (\ref{RDmodel0}),
we obtain the time-independent equations
\begin{subequations}
\label{spike1}
\begin{align}
0&=\epsilon^2 \frac{d^2 P}{d x^2}+   \frac{P^2}{A} - P, \\
0&=-D \frac{d A}{d x}-J, \\
0&=-\frac{dJ}{dx}+\frac{1}{\epsilon} P^2-A.
\end{align}
\end{subequations}
Suppose that there are $N$ spikes at positions $x_1,\ldots,x_{N}$ with $P$ peaking at each point $x_j$ such that 
\begin{equation}
P'(x_j)=0,\quad A(x_j)=U,
\end{equation}
and $U$ independent of $j$. (In order to apply asymptotic analysis, the spikes are located away from the boundaries $x=0,1$.)

Inside the inner region of the $j$-th spike we introduce the stretched coordinate $y_j=\epsilon^{-1}(x-x_j)$, and set $\widetilde{P}(y_j)=P(x_j+\epsilon y_j)$, $\widetilde{A}(y_j)=A(x_j+\epsilon y_j)$ and $\widetilde{J}(y_j)=J(x_j+\epsilon y_j)$. Introduce the asymptotic expansions 
\[\widetilde{P}=\widetilde{P}_0+o(1),\quad \widetilde{A}=\widetilde{A}_{0}+ {\epsilon} \widetilde{A}_{1}+O(\epsilon),\quad \widetilde{J}=\widetilde{J}_{0}+o(1),\]
with
\begin{align}
\label{P0}
\frac{d^2\widetilde{P}_0}{dy_j^2}+   \frac{ \widetilde{P}_0^2}{\widetilde{A}_0} - \widetilde{P}_0=0,\quad -\infty <y_j<\infty,
\end{align}
\begin{align}
\label{A0}
\frac{d\widetilde{A}_{0}}{dy_j}=0,\quad -D\frac{d \widetilde{A}_{1}}{d y_j}-\widetilde{J}_0=0, \quad
-\frac{d\widetilde{J}_{0}}{dy_j}+ \widetilde{P}_0^2=0, 
\end{align}
and
\begin{equation}
\widetilde{P}_0'(0)=0,\ \widetilde{A}_0(0)=U,\quad \widetilde{A}_1(0)=0.
\end{equation}
We also require $\widetilde{P}_0\rightarrow 0$ as $|y_j|\rightarrow \infty$. The analysis of $\widetilde{P}_0$, $\widetilde{A_0}$ and $\widetilde{A_1}$ now proceeds in an identical fashion to the classical GM model \cite{Iron01}. (However, we also have the additional variable $J$,which we will return to below).
The first equation of (\ref{A0}) immediately implies that $\widetilde{A}_0(y_j)=U$ for all $-\infty<y_j<\infty$. In addition, using phase plane analysis it can be shown that equation (\ref{P0}) has the solution \cite{Iron01}
\begin{equation}
\label{py}
\widetilde{P}_0(y_j)=\frac{3}{2}U\mbox{sech}^2\left (y_j/2\right ):=Up(y_j).
\end{equation}
Therefore, integrating the equation for $\widetilde{J}_0$ yields
\begin{equation}
\label{jump1}
\left [\lim_{y_j\rightarrow \infty}\widetilde{J}_0-\lim_{y_j\rightarrow -\infty}\widetilde{J}_0\right ]=  \Gamma U^2,\quad \Gamma = \int_{-\infty}^{\infty} \widetilde{p}(y)^2dy=6.
\end{equation}
Finally, substituting for $\widetilde{J}_0$ using the second equation in (\ref{A0}) implies that
\begin{equation}
\label{jump2}
\left [\lim_{y_j\rightarrow \infty}\widetilde{A}_1'-\lim_{y_j\rightarrow -\infty}\widetilde{A}_1'\right ]=-\frac{  \Gamma U^2}{D}.
\end{equation}

\begin{figure}[b!]
\begin{center} \includegraphics[width=8cm]{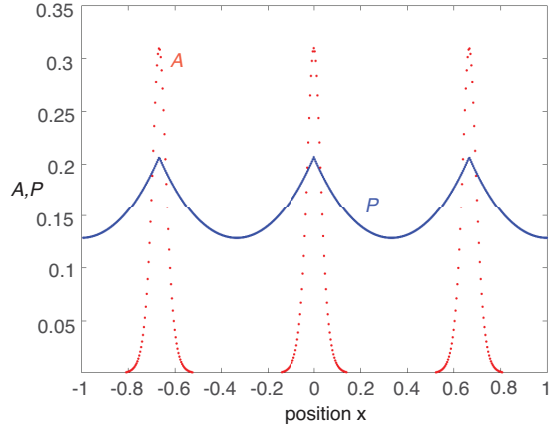} \end{center}
\caption{3-spike solution obtained from asymptotic analysis with $\epsilon=0.02$, $D=0.1$, $\rho_1=\mu_1=2\rho_2=\mu=1$. Plots of $A(x)$ and $P(x)$.}
\label{fig2}
\end{figure}

Equation (\ref{jump2}) acts as a jump condition for the first derivative of the solution for $A$ in the outer region. Since $P\sim 0$ in the outer region, we can write $A(x)=A_0(x)+O(\epsilon)$ with
\begin{equation}
D\frac{d^2A_0}{d x^2}-  A_0 =0,\quad x\in[-1,1]\backslash\{x_1,x_2,\ldots,x_N\},
\end{equation}
and $A_0$ continuous across $x_j$ but $A_0'$ discontinuous according to equation (\ref{jump2}). That is, 
\begin{equation}
D\frac{d^2A_0}{d x^2}-  A_0 =-  \Gamma U^2 \sum_{k=1}^N\delta(x-x_k),\quad x\in[-1,1],
\end{equation}
with $A_0'(x)=0$ at $x=0,1$. Introducing the Neumann Green's function
\begin{align}
D\frac{d^2G(x;x_k)}{d x^2}-  G(x;x_k)&=-\delta(x-x_k), \ \partial_xG(0;x_k)=0=\partial_xG(1;x_k),\\
 \int_{0}^{1}G(x;x_k)dx&=1,\nonumber
\end{align} 
we can write the outer solution as
\begin{equation}
\label{outer}
A_0(x)=\Gamma U^2 \sum_{j=1}^{N} G(x;x_j).
\end{equation}
The explicit form of the 1D Green's function is \cite{Iron01}
\begin{equation}
G(x;x_k)=\left \{ \begin{array}{cc} \calA_k\cosh\kappa(1+x)/\cosh\kappa (1+x_k) & \ -1<x<x_k\\
\calA_k\cosh\kappa(1-x)/\cosh\kappa (1-x_k) & \ x_k<x<1
\end{array}\right .,
\end{equation}
where
\begin{equation}
\calA_k=\kappa \left (\tanh\kappa(1-x_k)+\tanh\kappa (1+x_k)\right )^{-1},\quad \kappa=\sqrt{1/D}.
\end{equation}
In particular,
\begin{equation}
\label{G0}
\sum_{j=1}^{N} G(x;x_j)=\frac{1}{2\sqrt{D}\tanh(1/N\sqrt{D})}.
\end{equation}
It remains to determine the constant $U$. Setting $x=x_i$ in equation (\ref{outer}) yields the self-consistency condition
\begin{equation}
U=\Gamma U^2 \sum_{j=1}^{N}G(x_i;x_j).
\end{equation}
It follows that a symmetric $N$ spike solution will exist provided that the positions $x_1,\ldots,x_N$ are chosen such that the sum ${\mathcal G}=\sum_{j=0}^{N-1}G(x_i;x_j)$ is independent of $i$. 
It can be shown that this occurs for the equally spaced solution \cite{Iron01}
\begin{equation}
\label{eq11:equal}
x_j=-1+\frac{2j-1}{N},\quad j=1,\ldots,N.
\end{equation}
The non-trivial solution for the amplitude $U$ is then
\begin{equation}
\label{U}
U=\frac{1}{\Gamma \sum_{j=1}^{N}G(x_i;x_j)}.
\end{equation}

\begin{figure}[t!]
\begin{center} \includegraphics[width=12cm]{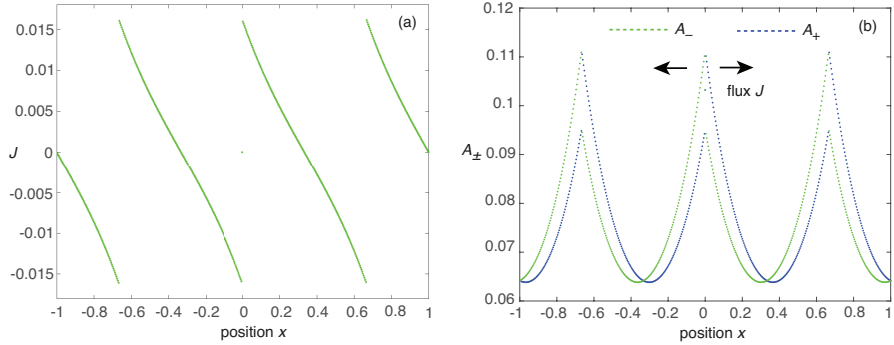} \end{center}
\caption{3-spike solution obtained from asymptotic analysis with $\epsilon=0.02$, $D=0.1$, $\rho_1=\mu_1=2\rho_2=\mu=1$. (a) Plot of flux $J(x)$. (c) Plots of $A_{\pm}(x)$.}
\label{fig3}
\end{figure}

So far everything has proceeded along the same lines as the analysis of spike solutions of the classical GM model. However, we also have the additional outer equation for $J(x)=J_0(x)+O(\epsilon)$:
\begin{equation}
J_0(x)=-D \frac{d A_0}{d x} \quad x\in[-1,1]\backslash\{x_1,x_2,\ldots,x_N\},
\end{equation}
with $A_0(x)$ given by equation (\ref{outer}) and $J_0$ automatically satisfying the jump condition (\ref{jump1}). Substituting for $A_0(x)$ gives
\begin{equation}
J_0(x)=-D \Gamma U^2\sum_{k=1}^{N}\partial_xG(x;x_k), \quad x\in[-1,1].
\end{equation}
In conclusion, in the limit $\epsilon \rightarrow 0$ the leading order asymptotic solutions for the unscaled concentrations $P$, $A$ and $J$ are of the form 
\begin{equation}
P(x)\sim Up([x-x_j]/\epsilon),\quad A(x)\sim  \Gamma U^2 \sum_{j=1}^{N} G(x;x_j),\quad J(x)\sim - D \Gamma U^2 \sum_{k=1}^{N}\partial_xG(x;x_k).
\end{equation}
In Figs. \ref{fig2} and \ref{fig3} we show an example of a 3-spike solution for $\epsilon=0.02$ and $D=0.1$. In particular, we see that the flux $J$ of the active component flows away from each peak such that the outer solutions for $A_+(x)$ and $A_-(x)$ are asymmetric.
\section{Linear stability analysis}

In order to investigate the stability of an equilibrium spike solution, we introduce the perturbations
\begin{equation}
P(x,t)=P(x)+\e^{\lambda t}\Phi(x),\quad A(x,t)=A(x)+\e^{\lambda t}\Psi(x),\quad J(x,t)=J(x)+\e^{\lambda t}\Lambda(x).
\end{equation}
Substituting into equations (\ref{RDmodel0}) and linearizing gives
\begin{subequations}
\label{RDlin}
\begin{align}
(1+\lambda) \Phi(x)&=\epsilon^2 \frac{d^2 \Phi(x)}{d x^2}+   2\frac{P(x)}{A(x)}\Phi(x)-\frac{P^2(x)}{A^2(x)}\Psi(x)  , \\
(1+\tau \lambda) \Psi(x)&=-\frac{d\Lambda(x)}{d x}+\frac{2P(x)}{\epsilon} \Phi(x)  ,\quad x\in [-1,1],\\
(1+\widehat{\tau} \lambda) \Lambda(x)&=-D\frac{d\Psi(x)}{d x} .
\end{align}
\end{subequations}
with 
\begin{equation}
\label{Dtau}
\tau=\frac{1}{\mu_2},\quad \widehat{\tau}= \frac{1}{2\alpha+\mu_2},\quad D=v^2\widehat{\tau}.
\end{equation}
These equations are supplemented by the no-flux boundary conditions
\begin{equation}
\Phi'(\pm1)=\Psi'(\pm 1)=\Lambda'(\pm 1)=0.
\end{equation} 
 Equation (\ref{Dtau}) implies that $0\leq \widehat{\tau}\leq \tau $, and varying $\widehat{\tau}$ from zero to $\tau$ is equivalent to going from the fast switching regime $\alpha \rightarrow \infty$ to the slow switching regime $\alpha \rightarrow 0$. In the following we investigate how the stability of the multi-spike solution depends on $\alpha$. 

Let us first consider the diffusion limit $\alpha \rightarrow \infty,v\rightarrow \infty$ such that $D$ is fixed. Eliminating $\Lambda(x)$ gives
\begin{subequations}
\label{RDlin2}
\begin{align}
(1+ \lambda) \Phi(x)&=\epsilon^2 \frac{d^2 \Phi(x)}{d x^2}+   2\frac{P(x)}{A(x)}\Phi(x)-\frac{P^2(x)}{A^2(x)}\Psi(x)  , \\
(1+\tau \lambda) \Psi(x)&=D\frac{d^2\Psi(x)}{d x^2}+\frac{2P(x)}{\epsilon} \Phi(x) ,\quad x\in [-1,1].
\end{align}
\end{subequations}
This is identical to the eigenvalue problem for the classical GM model studied in \cite{Iron01,Wei03}. One finds that there are two sets of eigenvalues: large or $O(1)$ eigenvalues that are bounded away from zero as $\epsilon \rightarrow 0$; small eigenvalues that approach zero as $O(\epsilon^2)$. The source of theses eigenvalues can be understood by considering the so-called shadow GM system in the limit $D\rightarrow \infty$ with $\tau=0$. The linear operator of the latter has $N$ exponentially small eigenvalues that reflect the near translation invariance of the system and the existence of exponentially weak interactions between neighboring spikes and between spikes and the boundary. These become the small eigenvalues for finite $D$ is finite. The linear operator of the shadow system also has exactly one positive eigenvalue in the vicinity of each spike, which implies that an $N$-spike solution is unstable on an $O(1)$ time-scale. However, as $D$ is decreased from infinity, the $O(1)$ eigenvalue near each spike moves into the left-half plane thus stabilizing the solution in the long-time limit. For $D<\infty$. $N\geq 2$, and $\tau \geq 0$, the following results hold (see propositions 4 and 11 of \cite{Iron01} and proposition 5.3 of \cite{Wei03}).
\smallskip

\noindent{\bf Small eigenvalues.} Consider a symmetric $N$-spike solution and the associated linear operator (\ref{RDlin2}) with $\tau=0$. The  $O(\epsilon^2)$ eigenvalues are negative only when $D<D_N^*$, where
\begin{equation}
D_N^*=\frac{1}{[N\ln(1+\sqrt{2})]^2}.
\end{equation}
There are $N-1$ small positive eigenvalues when $D>D_N^*$ and $\lambda=0$ is an eigenvalue of algebraic multiplicity $N-1$ when $D=D_N^*$. Moreover, $D_N^*$ is a decreasing function of $N$, with $D_2^*\approx 0.32$, $D_3^*\approx 0.14$ and $D_4^*\approx 0.08$, for example. These results persist to leading order when $\tau=O(1)$.
\smallskip

\noindent{\bf Large eigenvalues.}  Let $\lambda_0$ be the $O(1)$ eigenvalue with the largest real part for $\tau=0$. Then $\mbox{Re}(\lambda_0)<0$ when
\begin{equation}
\label{DN}
D<D_N\equiv \frac{1}{\Theta_N^2},\quad \Theta_N=\frac{N}{2}\ln \left [2+\cos(\pi/N)+\sqrt{\left (2+\cos(\pi/N))^2-1\right )}\right ].
\end{equation}
On the other hand, $\mbox{Re}(\lambda_0)>0$ when $D>D_N$. Note that $D_N>D_N^*$ and $D_N$ is also a decreasing function of $N$ with $D_1=\infty$, $D_2\approx 0.58$, $D_3\approx 0.18$ and $D_4\approx 0.09$, for example. It follows that there exists a parameter regime $D_N^*<D<D_N$ where an $N$-spike solution is stable on an $O(1)$ time-scale but unstable in the long-time limit. Finally, if $\tau=O(1)$ then an $N$-spike solution is unstable for any $\tau\geq 0$ when $D>D_N$. For $D>D_N$ and $\tau\geq 0$, there are eigenvalues that are real and positive, and the resulting instability involves spikes being annihilated in finite time. On the other hand, for $0<D<D_N$, an $N$-spike solution is stable with respect to the $O(1)$ eigenvalues provided that $0\leq \tau \leq \tau_N(D)$ for some critical time constant $\tau_N$ (which can be determined numerically). When $\tau$ crosses the critical value $\tau_N(D)$, a synchronous oscillatory instability of the spike amplitudes is induced.
\smallskip

\subsection{Derivation of nonlocal eigenvalue problem}

Now suppose that $v$ and $\alpha$ are both finite. Since the left-hand sides of equations (\ref{RDlin}b,c) involve the factors $(1+\tau \lambda)$ and $(1+\widehat{\tau} \lambda)$, respectively, it follows that the leading order results for the small eigenvalues carry over. Therefore, we can focus on the large eigenvalues. Equations (\ref{RDlin}) reduce to the pair of equations
\begin{subequations}
\label{RDlin3}
\begin{align}
(1+ \lambda) \Phi(x)&=\epsilon^2 \frac{d^2 \Phi(x)}{d x^2}+   2\frac{P(x)}{A(x)}\Phi(x)-\frac{P^2(x)}{A^2(x)}\Psi(x)  , \\
(1+\tau \lambda) \Psi(x)&=D_{\lambda}\frac{d^2\Psi(x)}{d x^2}+\frac{2P(x)}{\epsilon} \Phi(x)  ,\quad x\in [-1,1],
\end{align}
\end{subequations}
with $\Phi'(\pm1)=\Psi'(\pm 1)=0$ and a $\lambda$-dependent diffusivity
\begin{equation}
D_{\lambda}=\frac{D}{1+\widehat{\tau}\lambda}. 
\end{equation}
We also assume that $N\geq 2$ - the analysis of single spike solutions is slightly different \cite{Wei03}. In order to determine the $\alpha$-dependence of the eigenvalues for fixed $v^2$, it is necessary to consider the construction and analysis of the nonlocal eigenvalue problem developed in Ref. \cite{Wei03}.
Following along analogous lines to Refs. \cite{Iron01,Wei03}, we look for a localized eigenfunction of $\Phi$ in the form
\begin{equation}
\label{eqphi}
\Phi(x)\sim \sum_{k=1}^{N}c_k\phi[(x-x_k)/\epsilon).
\end{equation}
Since $\Phi$ is localized to the inner region of each spike, it follows that it can be treated as a sum of Dirac delta functions in the equation for $\Psi(x)$:
\begin{align}
D_{\lambda}\frac{d^2\Psi(x)}{d x^2}-(1+\tau \lambda) \Psi(x)=-2U\int_{-\infty}^{\infty}p(y)\phi(y)dy \sum_{k=1}^{N}c_k\delta(x-x_k).
\end{align}
Here $p(y)$ is defined in equation (\ref{py}). The solution for $\Psi(x)$ can be expressed in terms of the Neumann Green's function
\begin{align}
\frac{d^2G_{\lambda}(x;x_k)}{d x^2}-  \theta_{\lambda}G_{\lambda}(x;x_k)&=-\delta(x-x_k), \ \partial_xG_{\lambda}(0;x_k)=0=\partial_xG_{\lambda}(1;x_k),
\end{align} 
where 
\begin{equation}
\label{thet}
\theta_{\lambda}=\left [(1+\lambda \tau)/D_{\lambda}\right]^{1/2}=\sqrt{(1+\lambda \tau)(1+\lambda \widehat{\tau})/D}.
\end{equation}
That is,
\begin{equation}
\label{outer2}
\Psi(x)=\frac{\Omega}{D_{\lambda}}\sum_{k=1}^{N}c_kG_{\lambda}(x;x_k),
\end{equation}
with
\begin{equation}
\Omega=2U \int_{-\infty}^{\infty}p(y)\phi(y)dy.
\end{equation}
The explicit form of the 1D Green's function is \cite{Iron01}
\begin{equation}
G_{\lambda}(x;x_k)=\left \{ \begin{array}{cc} \calC_k\cosh\sqrt{\theta_{\lambda}}(1+x)/\cosh\sqrt{\theta_{\lambda}} (1+x_k) & \ -1<x<x_k\\
\calC_k\cosh\sqrt{\theta_{\lambda}}(1-x)/\cosh\sqrt{\theta_{\lambda}} (1-x_k) & \ x_k<x<1
\end{array}\right .,
\end{equation}
where
\begin{equation}
\calC_k=\frac{1}{\sqrt{\theta_{\lambda}}} \left (\tanh\sqrt{\theta_{\lambda}}(1-x_k)+\tanh \sqrt{\theta_{\lambda} }(1+x_k)\right )^{-1}.
\end{equation}

Substituting equations (\ref{eqphi}) and (\ref{outer2}) into equation (\ref{RDlin3}), using the fact that $A(x)=U+O(\epsilon)$ when $|x-x_j|=O(\epsilon)$, and expressing $\Phi$ in terms of stretched variables gives
\begin{align}
c_j\left ( \frac{d^2 \phi(y)}{d y^2}- \phi(y)+   2p(y)\phi(y)\right )- p^2(y) \frac{\Omega}{D_{\lambda}}\sum_{k=1}^{N}c_kG_{\lambda}(x_j;x_k) &=c_j \lambda \phi(y)
\end{align}
for $ -\infty<y<\infty$.
The eigenvalue problem is the same for each $j$ when $c_1,\ldots,c_{N}$ are the components of the eigenvector ${\bf c}$ of the matrix equation $\calG_{\lambda}{\bf c}=\alpha(\lambda) {\bf c}$, where $[\calG_{\lambda}]_{ij}=G_{\lambda}(x_i;x_j)$. We thus obtain the nonlocal eigenvalue problem
\begin{align}
\left ( \frac{d^2 \phi(y)}{d y^2}- \phi(y)+   2p(y)\phi(y)\right )- \frac{2\alpha(\lambda) p^2(y) }{ D_{\lambda}\sum_{j=1}^{N}G(x_i;x_j)}\frac{\int_{-\infty}^{\infty}p(y)\phi(y)dy}{\int_{-\infty}^{\infty}p^2(y)dy}&=  \lambda \phi(y)  
\end{align}
for $-\infty<y<\infty$, and we have substituted for $U$ using equation (\ref{U}). We also require $\phi\rightarrow 0$ as $|y|\rightarrow 0$.
The next step is to calculate the eigenvalues $\alpha(\lambda)$, which was carried out for the classical GM model in \cite{Iron01,Wei03}. The basic idea is to solve equation
(\ref{RDlin3}b) for $\Psi(x_n)$, $n=1,\ldots,N$, after rewriting it as
\begin{subequations}
\label{eqpsi}
\begin{align}
&D_{\lambda}\frac{d^2\Psi(x)}{d x^2}-(1+\tau \lambda) \Psi(x)=0,\quad x_{n-1}<x<x_n,\quad n=1,\ldots N+1,\\
&[\Psi]_n=0,\quad D_{\lambda} [\Psi'] =-\Omega c_n,\quad n=1,\ldots,N; \quad \Psi'(\pm 1)=0,
\end{align}
\end{subequations}
with $x_{0}=-1$ and $x_{N+1}=1$. One thus obtains the matrix equation 
\begin{equation}
{\mathcal B}{\bf a}=\frac{\Omega}{\sqrt{(1+\tau \lambda)D_{\lambda}}}{\bf c},
\end{equation} 
where ${\bf a}$ is the $N$-vector with components $\Psi(x_j)$ and ${\mathcal B}$ is a tridiagonal matrix \cite{Wei03}
\begin{equation}
{\mathcal B}=\left (\begin{array}{ccccccc} d_{\lambda} &f_{\lambda} & 0&\cdots &0& 0&0\\
f_{\lambda} &e_{\lambda} &f_{\lambda} &\cdots & 0 & 0 & 0\\
0 &f_{\lambda} &e_{\lambda} &\ddots & 0 & 0 & 0\\
\vdots &\vdots &\ddots &\ddots&\ddots&\vdots &\vdots\\
0&0&0&\ddots &e_{\lambda} &f_{\lambda}&0\\
0&0&0& \cdots &f_{\lambda}&e_{\lambda} &f_{\lambda}\\
0&0&0& \cdots &0& f_{\lambda}&d_{\lambda} 
\end{array}\right )
\end{equation}
with non-zero components
\begin{equation}
d_{\lambda}=\mbox{coth}(2\theta_{\lambda}/N)+\tanh(\theta_{\lambda}/N),\ e_{\lambda}=2\mbox{coth}(2\theta_{\lambda}/N),\ f_{\lambda}=-\mbox{csch}(2\theta_{\lambda}/N).
\end{equation}
Comparison with the solution (\ref{outer2}) for $x=x_i$ implies that
\begin{equation}
\frac{1}{D_{\lambda}}\calG=\frac{1}{\sqrt{(1+\tau \lambda)D_{\lambda}}}{\mathcal B}^{-1} .
\end{equation}
Hence, $\alpha(\lambda)= D_{\lambda}[(1+\tau \lambda)D_{\lambda}]^{-1/2} \kappa^{-1}(\lambda)$ where $\kappa(\lambda)$ is an eigenvalue of ${\mathcal B}$. The eigenvalues of ${\mathcal B}$ were calculated in \cite{Iron01,Wei03} for a given $\theta_{\lambda}$. We thus find that the eigenvalues of $\calG$ are
\begin{equation}
\alpha_j(\lambda)=\frac{D_{\lambda}}{2\sqrt{(1+\tau \lambda)D_{\lambda}}}\left [\mbox{coth}(2\theta_{\lambda}/N)-\mbox{csch}(2\theta_{\lambda}/N)\cos(\pi(j-1)/N)\right ]^{-1}.
\end{equation}

In summary, the $O(1)$ eigenvalues satisfy the set of nonlocal eigenvalue problems
\begin{align}
\label{spec}
\L_0\phi(y)-\chi_j(\lambda) p^2(y)\frac{\int_{-\infty}^{\infty}p(y)\phi(y)dy}{\int_{-\infty}^{\infty}p^2(y)dy}&=  \lambda \phi(y)  ,\ -\infty <y<\infty,\quad 
\end{align}
and $\phi \rightarrow 0\mbox{ as } |y|\rightarrow \infty$ for $j=0,\ldots,N-1$, with the linear operator $\L_0$ and factor $\chi_j(y)$ given by
\begin{align}
\label{spec0}
\L_0\phi(\lambda)&=\frac{d^2 \phi(y)}{d y^2}- \phi(y)+   2p(y)\phi(y),\\ 
\chi_j(y) &=\frac{2}{\sqrt{1+\tau \lambda}}
\tanh(1/N\sqrt{D}) \\
&\quad \times \left [\mbox{coth}(2\theta_{\lambda}/N)-\mbox{csch}(2\theta_{\lambda}/N)\cos(\pi(j-1)/N)\right ]^{-1}.\nonumber\end{align}
There are two sources of $\alpha$-dependence in the nonlocal eigenvalue equation (\ref{spec}): the term $\tanh(1/\sqrt{N}D)$ with $D=v^2\widehat{\tau}$ and the phase $\theta_{\lambda}$ satisfying equation (\ref{thet}).

\subsection{Stability conditions}

We now use equation (\ref{spec}) to derive stability conditions for the $O(1)$ eigenvalues. First, it is convenient to rewrite the nonlocal equation along the lines of \cite{Wei03}. Let $\psi(y)$ be the solution to the equation
\begin{equation}
\label{spec2}
\L_0\psi(y)\equiv \frac{d^2\psi}{dy^2}-\psi(y)+2p(y)\psi(y)=\lambda \psi(y)+p^2(y); \ \psi \rightarrow 0 \mbox{ as } |y|\rightarrow \infty.
\end{equation}
That is,
\begin{equation}
\psi=(\L_0-\lambda)^{-1}p^2.
\end{equation}
The eigenfunction satisfying equation (\ref{spec2}) can then be expressed as
\begin{equation}
\phi(y)=\chi_j(\lambda)\psi(y) \Upsilon,\quad \Upsilon =\frac{\int_{-\infty}^{\infty}p(y)\phi(y)dy}{\int_{-\infty}^{\infty}p^2(y)dy}.
\end{equation}
Multiplying both sides of the equation relating $\phi$ to $\psi $ by $p(y)$ and integrating with respect to $y$ yields a transcendental equation for the eigenvalues $\lambda$:
\begin{equation}
\label{g}
g_j(\lambda)\equiv C_j(\lambda)-f(\lambda)=0,\quad f(\lambda)\equiv \frac{\int_{-\infty}^{\infty}p(y)[\L_0-\lambda]^{-1}p^2(y)dy}{\int_{-\infty}^{\infty}p^2(y)dy},
\end{equation}
where
\begin{align}
\label{Cj}
C_j(\lambda)=\frac{1}{\chi_j(\lambda)}=\frac{\sqrt{1+\tau \lambda}}{2\tanh(1/N\sqrt{D}) }\left [\tanh(\theta_{\lambda}/N)+\frac{1-\cos(\pi(j-1)/N)}{\sinh(2\theta_{\lambda}/N)}\right ].
\end{align}
We have used some identities for hyperbolic functions. We wish to determine how the stability of an $N$-spike solution depends on $\alpha$. We begin by exploring conditions for the existence of positive real eigenvalues.

First consider the classical GM model, which is obtained in the limit $\alpha\rightarrow \infty$ and $v^2\rightarrow \infty$ such that $D$ is fixed and $\theta_{\lambda}=\sqrt{(1+\tau\lambda)/D}$.  The following results hold for real $\lambda$ \cite{Wei03}:
\medskip

\noindent (i)  The function $f(\lambda)$ has the asymptotic behavior
\begin{equation}
f(\lambda)\sim 1+\frac{3}{4}\lambda_R+\kappa_c\lambda^2+O(\lambda^3)\mbox{ as } \lambda\rightarrow 0; \quad f(\lambda)\rightarrow \infty \mbox{ as } \lambda\rightarrow \nu_0^-,
\end{equation}
where $\kappa_c>0$ and $\nu_0$ is the principal eigenvalue of the linear operator $\L_0$. Moreover, $f$ is a strictly monotonically increasing convex function for all $0<\lambda <\nu_0$, that is, $f'(\lambda)>0$ and $f''(\lambda)>0$; $f(\lambda)<0$ for $\lambda > \nu_0$.
\medskip

\noindent (ii)   For any fixed $D>0$ and $\tau >0$, each $C_{0,j}(\lambda)$, $j=0,\ldots,N-1$, is a strictly monotonically increasing concave positive function for all $\lambda \geq 0$. That is, $C_{j}(\lambda)>0$, $C_{j}'(\lambda)>0$, and $C_{j}''(\lambda)<0$. Moreover, $C_{j}'(\lambda)=O(\sqrt{\tau}) $ as $\tau \rightarrow \infty$.
\medskip

\noindent (iii)   The functions $C_j(\lambda)$ are ordered such that
\begin{equation}
C_{N}(\lambda)>C_{N-1}(\lambda)>\ldots > C_{1}(\lambda),\quad C_{N}'(\lambda)<C_{N-1}'(\lambda)>\ldots < C_{1}'(\lambda).
\end{equation}

\noindent (iv)   Set $B_j(D)=C_j(0)$ for $j=2,\ldots,N$. Here $B_j(D)$ is a strictly monotonically increasing function of $D$ for $D>0$ and is independent of $\tau$. In addition $B_j(D)<1$ for $0<D<D_N$ with $B_j(D_N)=1$ and $D_N$ given by equation (\ref{DN}).
\medskip

\begin{figure}[t!]
\begin{center} \includegraphics[width=12cm]{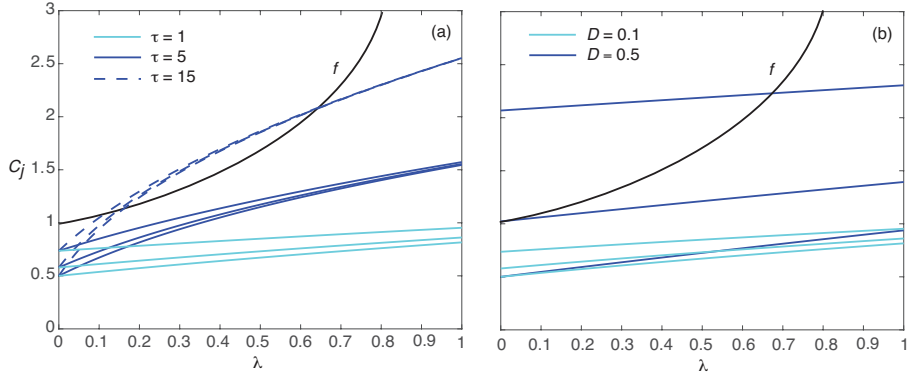} \end{center}
\caption{Plots of the functions $C_j(\lambda)$, $j=1,2,3$, for a 3-spike solution of the classical GM model. (a)  $D=0.1<D_3$, and various $\tau$. (b) $\tau=0.1$ and different diffusivities $D$. For each triplet, the lowest curve is for $C_1$ and the highest curve is for $C_3$. The black convex curve is the function $f(\lambda)$.}
\label{fig4}
\end{figure}

\noindent It follows from (i)-(iv) that if $D<D_N$ and $\tau$ is sufficiently small (so that $C_{j}'(\lambda) $ is sufficiently small), the curves $f(\lambda)$ and $C_j(\lambda)$ will not intersect for any $\lambda \geq 0$. That is, the $N$-spike solution is stable with respect to eigenmodes corresponding to real eigenvalues. The basic picture is illustrated in Fig. \ref{fig4}(a) for a 3-spike solution and $D<D_3$. One finds that there are no points of intersection for $\tau <\tau_c$ with $\tau_c=O(10)$. On the other hand, for large $\tau$, there are two real eigenvalues for each $j=1,2,3$. It is clear that these eigenvalues do not appear by crossing the origin. Indeed, as highlighted in \cite{Wei03}, instabilities induced by increasing $\tau$ for fixed $D$ can only occur via a Hopf bifurcation. On the other hand, instabilities induced by increasing $D$ for fixed $\tau$ can occur via a real eigenvalue entering the right half-plane, see Fig. \ref{fig4}(b).  

\begin{figure}[t!]
\begin{center} \includegraphics[width=12cm]{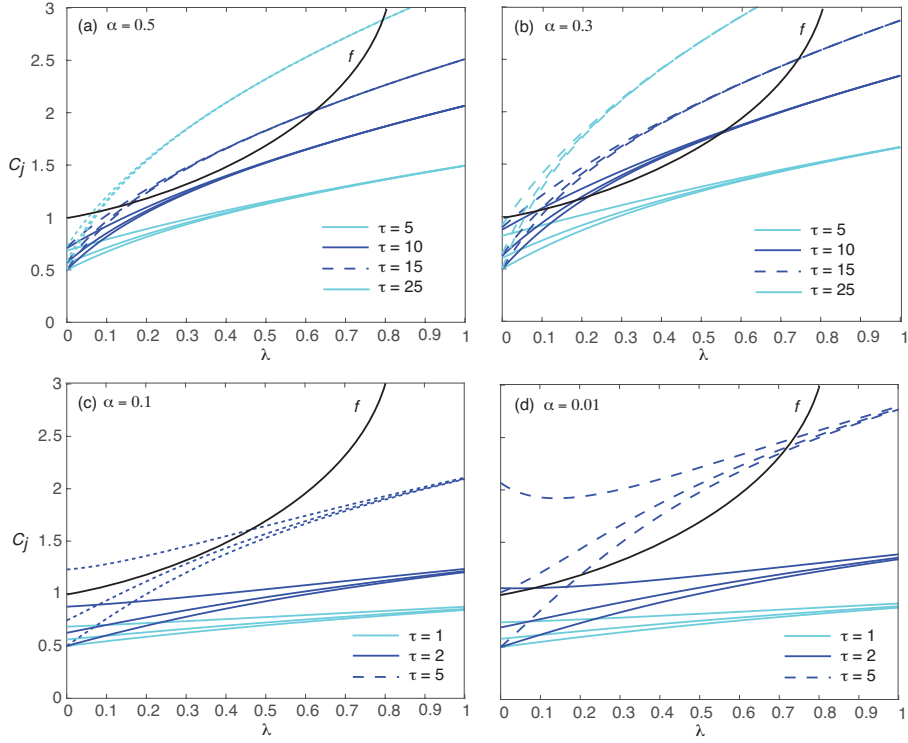} \end{center}
\caption{Plots of the functions $C_j(\lambda)$, $j=1,2,3$, for a 3-spike solution of the hybrid reaction-transport model. Here $v^2=0.1$ and $\widehat{\tau}=\tau/(1+2\tau \alpha)$. (a) $\alpha=0.5$; (b) $\alpha=0.3$; (c) $\alpha=0.1$ and (d) $\alpha=0.01$. In each case, the lowest curve is $C_1$ and the highest curve is $C_3$.The black convex curve is the function $f(\lambda)$.}
\label{fig5}
\end{figure}

How does this story change for finite $\alpha$ and $v^2$? We will explore this issue by fixing the speed $v$, and varying the switching rate $\alpha$. This has two distinct effects. First, it modifies the effective diffusivity $D=v^2\widehat{\tau}$ and, second, it alters the phase $\theta_{\lambda}$. For sufficiently large values of $\alpha$, the time constant $\widehat{\tau}$ is only weakly dependent on $\tau$ so increasing the latter results in the appearance of pairs of positive real eigenvalues that do not cross the origin, indicative of a Hopf instability, see Fig. 5(a,b). On the other hand, in the limit $\alpha\rightarrow 0$, we have $\widehat{\tau}\approx \tau$ and the dominant effect of raising $\tau$ is to increase the effective diffusivity. This implies that the resulting instability involves a real eigenvalue crossing the origin, as illustrated in Fig. \ref{fig5}(c,d). A necessary condition for no positive real eigenvalues is $v^2\widehat{\tau}<D_N$, that is, $[v^2-2\alpha D_N]\tau <D_n $. The latter inequality is always satisfied if $\alpha>\alpha_c$ and holds for $\tau<\tau_c(\alpha)$ when $\alpha <\alpha_c$, where
\begin{equation}
\alpha_c =\frac{v^2}{2D_N};\quad \tau_c(\alpha)=\frac{D_N}{v^2-2\alpha D_N} \mbox{ for } \alpha <\alpha_c.
\end{equation}
In the case $N=3$, we have $D_3\approx 0.1$ so that $\alpha_c\approx 0.28$ when $v^2=0.1$. In cases where $v^2\widehat{\tau}<D_N$, we still require $\tau$ to be sufficiently small so that there are no points of intersection with $f(\lambda)$.

The next step is to look for pure imaginary eigenvalues. We begin by separating equation (\ref{g}) into real and imaginary parts:
\begin{equation}
g_j=g_{j,R}+ig_{j,I},\quad f=f_R+if_I,\quad \lambda=\lambda_R+i\lambda_I.
\end{equation}
If we now set $\lambda_R=0$, the eigenvalues along the imaginary axis are the solutions of the following pair of equations
\begin{equation}
\widetilde{g}_{j,R}(\lambda_I)\equiv \widetilde{C}_{j,R}(\lambda_I)-\widetilde{f}_{R}(\lambda_I),\quad \widetilde{g}_{j,I}(\lambda_I)\equiv \widetilde{C}_{j,I}(\lambda_I)-\widetilde{f}_{I}(\lambda_I),
\end{equation}
with $\widetilde{C}_{j,R}(\lambda_I)=\mbox{Re}[C_j(i\lambda_I)]$, $\widetilde{C}_{j,L}(\lambda_I)=\mbox{Im}[C_j(i\lambda_I)]$, and
\begin{align*}
\widetilde{f}_{R}(\lambda_I)&=\mbox{Re}[f(i\lambda_I)]=\frac{\int_{-\infty}^{\infty}p(y)\L_0[\L_0^2+\lambda_I^2]^{-1}p^2(y)dy}{\int_{-\infty}^{\infty} \ p^2(y)dy}\\
\widetilde{f}_{I}(\lambda_I)&=\mbox{Im}[f(i\lambda_I)]=\frac{\lambda_I \int_{-\infty}^{\infty}p(y) [\L_0^2+\lambda_I^2]^{-1}p^2(y)dy}{\int_{-\infty}^{\infty} \ p^2(y)dy}.
\end{align*}
Without loss of generality, we can take $\lambda_I \geq 0$. The following results for $\widetilde{f}_{R}(\lambda_I)$ and $\widetilde{f}_{I}(\lambda_I)$ hold \cite{Wei03}:
\medskip

\begin{figure}[t!]
\begin{center} \includegraphics[width=12cm]{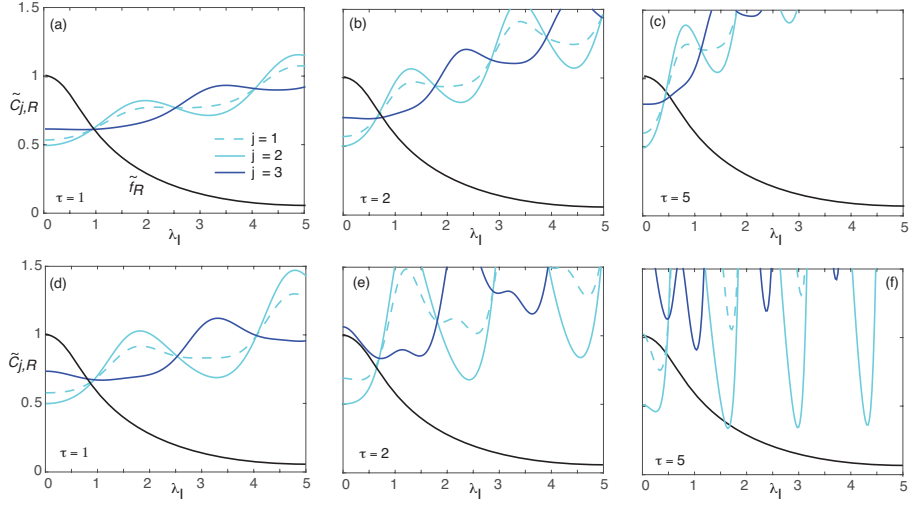} \end{center}
\caption{Plot of the functions $\widetilde{C}_{j,R}(\lambda_I)$, $j=1,2,3$, for a 3-spike solution of the hybrid reaction-transport model. Here $v^2=0.1$ and $\widehat{\tau}=\tau/(1+2\tau \alpha)$ with $\tau=1,2,5$. (a-c) $\alpha=0.3$; (d-f) $\alpha=0.01$. The black curve is the function $\widetilde{f}_R(\lambda_I)$.}
\label{fig6}
\end{figure}

 \noindent (a)  The function $\widetilde{f}_{R}$ has the asymptotic behavior
\begin{equation}
\widetilde{f}_{R}(\lambda_I) \sim 1-\kappa_c \lambda_I^2+O(\lambda_I^4) \mbox{ as } \lambda_I\rightarrow 0; \quad \widetilde{f}_{R}(\lambda_I)\rightarrow O(\lambda_I^{-2}) \mbox{ as } \lambda_I\rightarrow \infty,
\end{equation}
where $\kappa_c>0$. Moreover, $f$ is a strictly monotonically decreasing  function for all $0<\lambda_I <\infty$, that is, $\widetilde{f}_{R}'(\lambda_I)<0$.
\medskip

 \noindent (b)  The function $\widetilde{f}_{I}$ has the asymptotic behavior
\begin{equation}
\widetilde{f}_{I}(\lambda_I) \sim \frac{3}{4} \lambda_I+O(\lambda_I^3) \mbox{ as } \lambda_I\rightarrow 0; \quad \widetilde{f}_{I}(\lambda_I)\rightarrow O(\lambda_I^{-1}) \mbox{ as } \lambda_I\rightarrow \infty,
\end{equation}
and $\widetilde{f}_{I}(\lambda_I)>0$ for all $0<\lambda_I <\infty$.
\medskip

\noindent Again we begin by considering the classical GM model in the limit $\alpha\rightarrow infty$ and $v\rightarrow \infty$ with $D$ fixed. Note that $\widetilde{C}_{j,R}(0)=C_j(0)=B_j(D)$ for all $\lambda_I$. Hence, if $D<D_N$ then $\widetilde{C}_{j,R}(0)<\widetilde{f}_{R}(0) $. Moreover, if $\tau>0$ then $\widetilde{C}_{j,R}(\lambda_I)$ is an increasing function of $\lambda_I$ \cite{Wei03}, which from result (a) means that there is a single intersection point with $\widehat{f}_{R} (\lambda_I)$ and thus a unique positive root $\lambda_I^*$ where $\widetilde{g}_{j,R}(\lambda_I^*)=0$. Moreover, from result (b) we have $\widetilde{f}_I(\lambda_I)>0$ for all $\lambda_I >0$, whereas $\widetilde{C}_{j,I} =O(\tau)$ as $\tau\rightarrow 0$. This implies that for sufficiently small $\tau$, $\widetilde{g}_{j,I}(\lambda_I^*) <0$ and there are no purely imaginary eigenvalues. 

Now suppose that $\alpha$ and $v$ are finite. Similar to the case of real eigenvalues, we wish to explore how $\widetilde{C}_{j,R} (\lambda_I)$ and $\widetilde{C}_{j,I} (\lambda_I)$ depend on the switching rate $\alpha$. Some example plots of $\widetilde{C}_{j,R} (\lambda_I)$ are shown in Fig. \ref{fig6} for the two cases $\alpha=0.3$ and $\alpha=0.01$. It is clear that $\widetilde{C}_{j,R} (\lambda_I)$ is no longer a monotonically increasing function of $\lambda_I$, even though $\widetilde{C}_{j,R} (\lambda_I)\rightarrow \infty$ as $\lambda_I \rightarrow \infty$.That is, there still exists at least one intersection point between $\widetilde{C}_{j,R} (\lambda_I)$ and $\widetilde{f}_{R} (\lambda_I)$, but it is no longer necessarily unique unless $\tau$ is sufficiently small; the latter condition ensures that the curves are only weakly non-monotonic. For larger values of $\tau$, one has to investigate the number of intersection points numerically. Although we cannot make any rigorous statements, our numerical studies suggest that there is a unique intersection point for all $\tau >0$ when $\alpha>\alpha_c$, as illustrated in Fig. \ref{fig6}(a-c). On the other hand, for $\alpha <\alpha_c$ and sufficiently large $\tau$, it is possible for $\widetilde{C}_{j,R} (\lambda_I)$ to intersect $\widetilde{f}_{R} (\lambda_I)$ more than once, see Fig. \ref{fig6}(d-f).

\begin{figure}[t!]
\begin{center} \includegraphics[width=12cm]{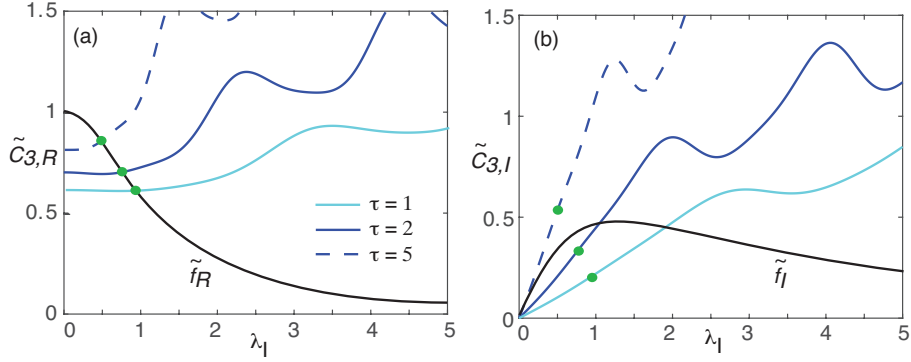} \end{center}
\caption{Plot of the functions (a) $\widetilde{C}_{3,R}(\lambda_I)$ and (b) $\widetilde{C}_{3,I}(\lambda_I)$ for a 3-spike solution of the hybrid reaction-transport model. Here $v^2=0.1$ and $\widehat{\tau}=\tau/(1+2\tau \alpha)$ with $\tau=1,2,5$ and $\alpha=0.3$. The black curves in (a) and (b) are the functions $\widetilde{f}_R(\lambda_I)$ and $\widetilde{f}_I(\lambda_I)$, respectively. Green dots indicate the values at $\lambda_I=\lambda_I^*$.}
\label{fig7}
\end{figure}

\begin{figure}[b!]
\begin{center} \includegraphics[width=12cm]{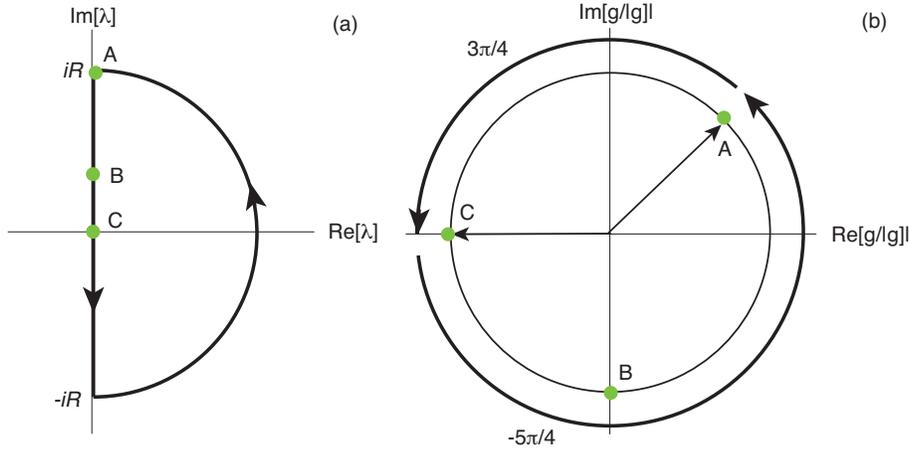} \end{center}
\caption{Winding number argument of \cite{Wei03}. (a) Counterclockwise path along the contour $\Gamma_R\cup \Gamma_I\cup \overline{\Gamma}_I$ in the complex $\lambda$-plane, where $\Gamma_R$ is the semi-circle of radius $R$, $\Gamma_I=[iR,0]$ and $\overline{\Gamma}_I=[0,-iR]$. (b) Change in $\arg g$ along $\Gamma_I$ (from A to C) in the limit $R\rightarrow \infty$. There exists a unique point $i\lambda^*$ on the positive imaginary axis where $\widetilde{g}_{j,R}(\lambda^*)=0$ (point B). If $\widetilde{g}_I(\lambda^*)<0$ then $\arg g(i\lambda^*)=-\pi/2$ and $\Delta \arg g = -5\pi/4$.}
\label{fig8}
\end{figure}

For the moment, take $\alpha>\alpha_c$ and assume that for the given $\tau$ there exists a unique point $i\lambda_I^*$ on the imaginary axis at which $\widetilde{g}_{j,R}(\lambda_I^*)=0$. It can be checked that if $\tau$ is sufficiently small then $\widetilde{g}_{j,I}(\lambda_I^*)<0$. This is illustrated in Fig. \ref{fig7}, where we plot $\widetilde{C}_{3,R}(\lambda_I)$ and $\widetilde{C}_{3,I}(\lambda_I)$ for $\alpha=0.3$ and various $\tau$. As $\tau$ increases the curve $\widetilde{C}_{3,R}(\lambda_I)$ becomes steeper in a neighborhood of the origin so that $\widetilde{g}_{j,I}(\lambda_I^*)$ switches sign. The condition $\widetilde{g}_{j,I}(\lambda_I^*)<0$ not only ensures that there are no pure imaginary eigenvalues but also implies that there are no complex-valued eigenvalues in the right complex half-plane. This follows from the winding number construction of \cite{Wei03}. First, recall the argument principle of complex analysis: if $F(x)$ is a meromorphic function inside and on some closed contour $C$ and $F$ has no zeros or poles on $C$, then
\begin{equation}
\frac{1}{2\pi i}\oint_C \frac{F'(z)}{F(z)}dz=\frac{1}{2\pi i}\oint_{F(C)}\frac{dw}{w }=Z-P,
\end{equation}
where $w=F(z)$, and $Z,P$ denote, respectively, the number of zeros (including multiplicity) and poles (including order) of $F(z)$ inside $C$. The second integral is the winding number of $F$. Consider the contour $C$ shown in Fig. 8(a), which consists of the semicircle $\Gamma_R $ given by $|\lambda|=R>0$ for $-\pi/2 \mbox{arg}\lambda \leq \pi/2$, and the section $-iR\leq \mbox{Im}[\lambda]\leq iR$ of the imaginary axis. Assuming that $\alpha$ is sufficiently large so that there are no pure imaginary eigenvalues (see above), we can let $R\rightarrow \infty$ and use the argument principle to determine the number of zeros of $g_j(\lambda)$ in the right-half plane. Since $C_j(\lambda)\sim \lambda^{1/2}$ as $|\lambda|\rightarrow \infty$ while $f(\lambda)\rightarrow 0$ as $|\lambda|\rightarrow \infty$, it follows that $\arg\lambda$ changes by an amount $\pi/2$ as $\Gamma_R$ is traversed counterclockwise. Using the fact that $g_j(\lambda)$ is analytic in the right-half plane, except at the simple pole $\lambda=\nu_0>0$ ($P=1$), it follows from the argument principle that the number $M$ of zeros of $g_j(\lambda)$ in the right half-plane is
\begin{equation}
\label{Z}
Z=\frac{5}{4}+\frac{1}{\pi}[\arg g_j]_{\Gamma_I},
\end{equation}
where $[\arg g_j]_{\Gamma_I}$ is the change in the argument of $g$ along the semi-infinite imaginary axis $\Gamma_I=i\lambda $, $0\leq \lambda <\infty$. Setting $\widetilde{C}_j(\lambda) = C_j(i\lambda)$ with $\lambda$ real, we have $\widetilde{C}_j(\lambda)\sim \sqrt{i\lambda}$ as $\lambda \rightarrow \infty$ so that $\arg g=\pi/4$ at the start of $\Gamma_I$. Moreover, $\arg g_j(0) = \pi$, since $g_j(0)
=C_j(0)-f(0)<0$. Given the unique point $i\lambda^*$ on the positive imaginary axis where $\widetilde{g}_{j,R}(\lambda^*)=0$. If we can show that $\widetilde{g}_{j,I}(\lambda^*)<0$, then $\arg g(i\lambda^*)=-\pi/2$ and the change of $\arg g$ along $\Gamma_I$ is $-5\pi/4$, see Fig. \ref{fig8}(b). Equation (\ref{Z}) then ensures that there are no eigenvalues in the right half-plane and the $N$-spike solution is stable. From result (b), we have $\widetilde{f}_{I}(\lambda)>0$ for all $0<\lambda <\infty$, whereas $\widetilde{C}_{j,I}(0)=0$ and $\widetilde{C}_{j,I}(\lambda)=O(\tau)$ as $\tau \rightarrow 0$ for finite $\lambda$. Hence, for $j=1,\ldots,N$, there exists a $\tau_0>0$ such that $\widetilde{g}_{j,I}(\lambda^*)<0$ and thus $Z=0$ for all $\tau$ satisfying $0\leq \tau<\tau_0$. 

\begin{figure}[t!]
\begin{center} \includegraphics[width=12cm]{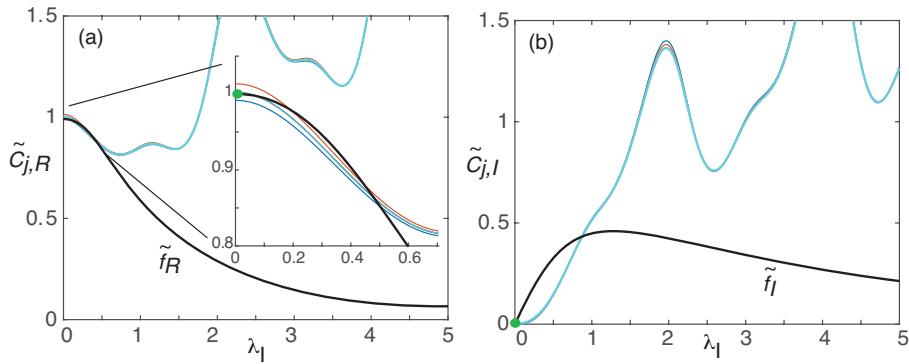} \end{center}
\caption{Plot of the functions (a) $\widetilde{C}_{3,R}(\lambda)$ and (b) $\widetilde{C}_{3,I}(\lambda)$ for a 3-spike solution of the hybrid reaction-transport model. Here $v^2=0.1$ and $\widehat{\tau}=\tau/(1+2\tau \alpha)$ with $\tau=2$. Solutions for three slightly different values of the switching rate are shown:  $\alpha=0.02,0.025,0.03$. As $\alpha$ increases, $\widetilde{C}_{3,R}(\lambda)$ crosses $\widetilde{f}_{R}(\lambda)$ at the origin (highlighted by the green dot in the inset), indicating that a real eigenvalue becomes positive. For a narrow range of $\alpha$ there exists a pair of intersection points between $\widetilde{C}_{3,R}(\lambda)$ and $\widetilde{f}_{R}(\lambda)$.}
\label{fig9}
\end{figure}

We have thus established that the hybrid reaction-transport model with GM kinetics supports a stable $N$-spike solution that is stable on $O(1)$ time-scales provided that $\alpha>\alpha_c$ and $0\leq \tau < \tau_0$ for some $\tau_0=\tau_0(\alpha)$. The first condition ensures that $D=v^2\widehat{\tau}<D_N$, with $D_N$ given by equation (\ref{DN}). When $\tau$  increases beyond $\tau_0$, a pair of complex conjugate eigenvalues crosses into the right half-plane, indicating the occurrence of a Hopf bifurcation. Finally, note that if $\alpha < \alpha_c$, then the same winding number argument can be used to show that an $N$-spike is still stable for sufficiently small $\tau$, but the nature of the instability as $\tau$ increases differs. This is due to the fact that $D$ crosses the critical value $D_N$, signaling that a real eigenvalue passes the origin into the right half-plane. An example of such a scenario is shown in Fig. \ref{fig9}.

\section{Conclusion}

In summary, we have extended the asymptotic analysis of the classical two-component GM model \cite{Iron01,Wei03} to a hybrid reaction-transport model with GM kinetics in order to derive conditions for the existence and stability of an $N$-spike solution. We exploited the fact that in the limit $\alpha\rightarrow \infty$ and $v\rightarrow \infty$ with $D=v^2/(\mu_2+2\alpha)$ fixed, the 3-component hybrid reaction-transport model reduces to the classical two-component GM reaction-diffusion model with effective inhibitor diffusivity $D$. Here $v$ is the speed of active transport, $\alpha$ is the switching rate and $\mu_2$ is the degradation rate of the inhibitor. Linearizing about a multi-spike solution resulted in a non-local eigenvalue problem that could be used to derive stability conditions as a function of the switching rate $\alpha$. In particular, we showed that there exists a critical switching rate $\alpha_c$ such that for $\alpha >\alpha_c$, the multi-spike solution tends to undergo a Hopf instability, whereas for $\alpha <\alpha_c$ the instability is due to one or more real eigenvalues becoming positive.
Although we focused our analysis on the particular hybrid model introduced within the context of synaptogenesis in {\em C. elegans} \cite{Brooks16,Brooks17}, it would be interesting in future work to explore a more general class of hybrid models. One generalization would be to consider alternative reaction schemes to GM kinetics, where it is not possible to reduce the steady-state equations to a two-component reaction-diffusion system by eliminating the flux $J=v(A_+-A_-)$. This would require modifying the construction of multi-spike solutions in section 3. A second modification would be to allow for biased active transport by taking the speeds of the left-moving and right-moving states to differ. Any effective reduction of the hybrid model would result in an advection-diffusion equation for the inhibitor. Third, from a mathematical perspective, there is no reason to identify the activator as the diffusing species and the inhibitor as the actively transported species. How would the results change if the transport modes of the activator and inhibitor were reversed?

The particular application of the hybrid model to larval development also raises a number of issues. For example, while the important role of CaMKII in regulating the delivery of GLR-1 to synapses along the  ventral cord of {\em C. elegans} is well known, the detailed mechanisms regarding their interactions are still unclear. Developing a more detailed biophysical model of CaMKII-GLR-1 coupling would be one way to explore more general reaction schemes in the hybrid model. This might also require the introduction of additional components, such as allowing for stationary states of GLR-1 and distinguishing between cytoplasmic and membrane-bound CaMKII. Another major aspect of {\em C. elegans} development is the insertion of new synaptic puncta as the larva grow, see Fig. \ref{fig1}. It would be interesting to model this as an example of spike insertion in a growing domain.

\vskip6pt

\enlargethispage{20pt}








\end{document}